\documentclass[10pt, conference]{IEEEtran}
\IEEEoverridecommandlockouts
\usepackage{cite}
\usepackage{amsmath,amssymb,amsfonts}
\usepackage{algorithmic}
\usepackage{graphicx}
\usepackage{textcomp}
\usepackage{xcolor}

\begin{document}

\title{MS2M: A message-based approach for\\live stateful microservices migration
}

\author{\IEEEauthorblockN{Hai Dinh-Tuan}
\IEEEauthorblockA{\textit{Service-Centric Networking} \\
\textit{Technische Universit\"at Berlin}\\
Berlin, Germany \\
hai.dinhtuan@tu-berlin.de}
\and
\IEEEauthorblockN{Felix Beierle}
\IEEEauthorblockA{\textit{Institute of Clinical Epidemiology and Biometry} \\
\textit{University of W\"urzburg}\\
W\"urzburg, Germany \\
felix.beierle@uni-wuerzburg.de}
}

\maketitle

\begin{abstract}
In the last few years, the proliferation of edge and cloud computing infrastructures as well as the increasing number of mobile devices has facilitated the emergence of many novel applications. However, that increase of complexities also creates novel challenges for service providers, for example, the efficient management of interdependent services during runtime. One strategy is to reallocate services dynamically by migrating them to suitable servers. However, not every microservice can be deployed as stateless instances, which leads to suboptimal performance of live migration techniques. In this work, we propose a novel live migration scheme focusing on stateful microservices in edge/cloud environments by utilizing the underlying messaging infrastructure to reconstruct the service's state. Not only can this approach be applied in various microservice deployment scenarios, experimental evaluation results also show a reduction of 19.92\% downtime compared to the stop-and-copy migration method.
\end{abstract}


\section{Introduction and Motivation}
In the last few years, microservice architectures have attracted attention from both academia and industry as a novel approach to software development. In fact, large cloud service providers such as Amazon or Netflix have widely adopted this new architecture to serve millions of users on a daily basis \cite{thones2015microservices}. In essence, this software design encourages the decomposition of monoliths into smaller, independently deployable units, each provides one or several services \cite{fowler2018}.

In the \textit{Twelve-Factor App methodology} \cite{wiggins2012twelve}, the authors have described an ideal scenario, in which all microservices should be designed and implemented as stateless instances. Since stateless services do not have the responsibility to manage state data internally, and no prior knowledge is needed, each request can be processed by any service instance. This enables applications to achieve a very high level of scalability and resiliency, because multiple service instances can be quickly deployed to handle peak loads. Similarly, failed service instances can also be replaced by launching other replacement instances. However, it is not always feasible to design a system based entirely on stateless services \cite{furda2017migrating}. Over the years, several workarounds have been proposed to delegate state data management to an external entity. For example, a service can adopt the \textit{client side state management technique} to store state information in the clients instead. Another option is to constantly save and load service states from an external database. However, in both cases, additional overheads caused by state data exchanges reduce the application's latency performance \cite{ingeno2018software} and generate more network loads. 

In distributed systems such as microservices, a reliable communication channel among services plays a crucial role to orchestrate services, especially in the \textit{event-driven} design, where all the events and state information are constantly exchanged among service instances. Therefore, this communication channel can be considered as the backbone of a microservice-based application, and in fact, some existing frameworks even require the developers to design the data communication before implementing the services \cite{dinh2020development}. Thus, in this paper, we investigate how to exploit the already mature communication infrastructure for microservices to solve the state management problem and develop an efficient migration scheme. 

This paper focuses on service migration, where the problem of state management is most prominently presented. In recent years, with the popularity of cloud-based services as well as the advent of new concepts such as osmotic computing \cite{villari2016osmotic}, multi-access edge computing \cite{tanaka2018multi}, service migration has been increasingly discussed in the literature as a way to not only achieve resource management goals (power management, resource load balancing) but also to improve application's flexibility, fault tolerance, etc. This paper describes a novel approach to service migration using the existing message exchange mechanisms in many microservices-based applications to support stateful microservices migration without adding a dedicated state management entity. We name our proposed approach MS2M, which is an acronym for \textit{Message-based stateful microservice migration}.

In this work, we implemented a prototype using a remake version of Flappy Bird -- one of the most iconic mobile games originally launched in 2013. This interactive game is chosen so that we can easily demonstrate how the migration process affects the game quality. We have modified the codebase with new components to evaluate our migration scheme. While the user interacts with the browser-based component, one cloud-based component is migrated to a cloud region with better latency performance. The goal is to deliver a seamless experience for the players, i.e., the game is not interrupted during the migration process.

The contribution of this work is therefore two-fold: (1) design and implementation of a novel technique to facilitate stateful microservices portability using the existing message exchange mechanism, and (2) a performance evaluation of a prototype performed in a typical cloud environment. 

In the remainder of this paper, Section II discusses the existing migration techniques that have been found in the literature, followed by a detailed description of the proposed concept for service migration using asynchronous messaging in Section III. Section IV briefly explains the evaluation setup, and the results are also presented. Finally, we discuss the results and conclude this work in Section V.

\section{Related work} 

\subsection{Service migration techniques}

Currently, microservices are usually encapsulated into virtual machines and containers to simplify the deployment and avoid interoperability issues \cite{pahl2016microservices}. Therefore, we focused on related migration techniques for those deployment technologies.

\begin{figure}
\includegraphics[width=\linewidth]{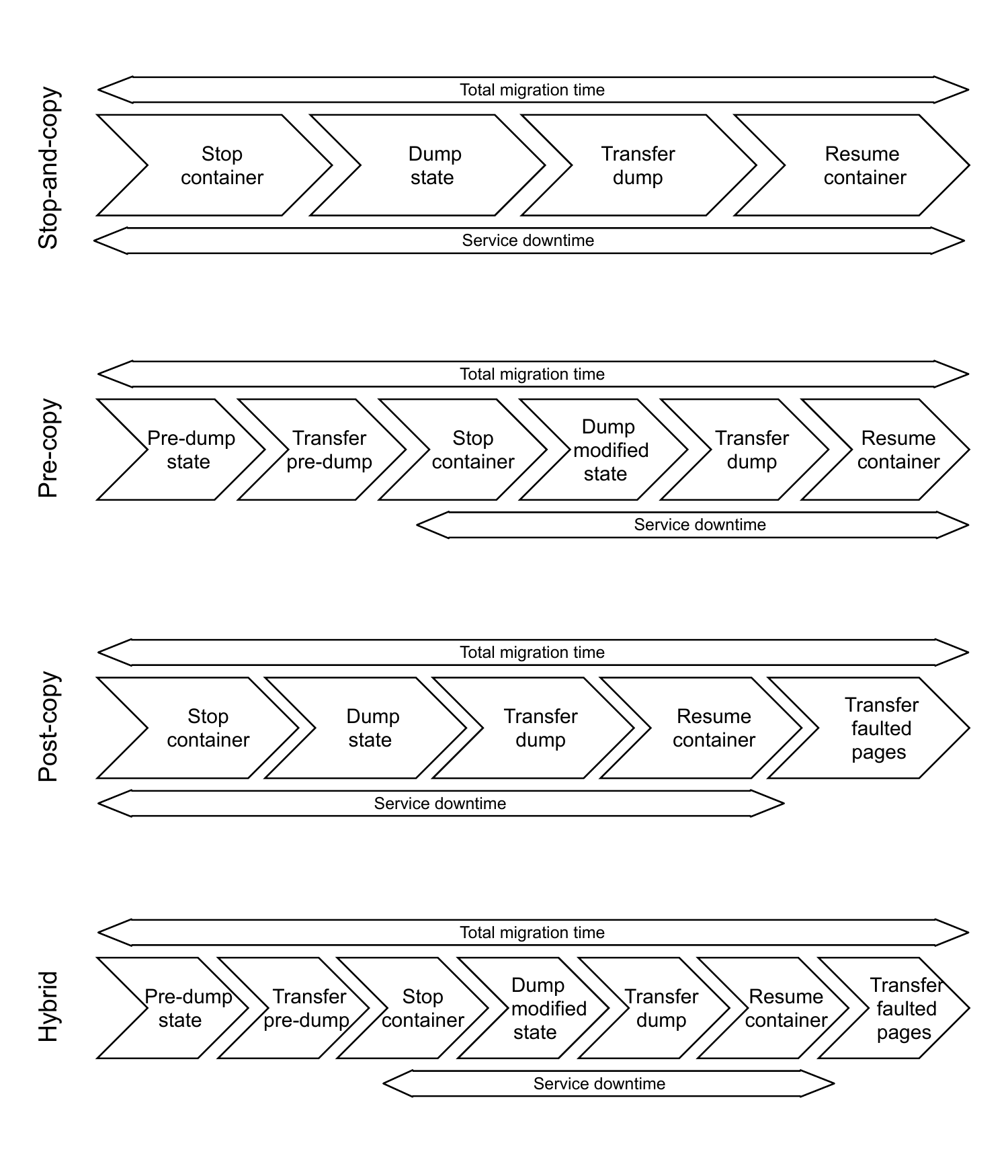}
    \caption{Overview of stop-and-copy, pre-copy, post-copy and hybrid migration techniques.}
    \label{fig:overview}
\end{figure}

As virtual machines have been widely utilized for a long time, virtual machines migration is a well in-depth analyzed topic with plenty of proposals \cite{basu2019learn, pal2019novel}. Virtual machines persist their states in virtual disks, and they are migrated across platforms using various tools, both proprietary like Hyper-V Live Migration or third-party tools. The simplest approach is \textit{cold migration} or \textit{stop-copy}, in which the execution is first stopped at the source, after the state is transferred and restored at the destination, the execution continues at the new location \cite{sapuntzakis2002optimizing, kozuch2002internet, whitaker2004constructing}. As a result of its simplicity, this approach minimizes the amount of data transfer between source and destination, while maintaining a small total migration time in comparison to more sophisticated techniques. However, both downtime and total migration time are proportional to the amount of data that need to be transferred. In addition, there is a downtime period where no client request can be processed. To minimize this downtime period, there are several live migration schemes proposed such as pre-copy migration \cite{theimer1985preemptable}, post-copy migration \cite{hines2009post}, or hybrid \cite{lei2017novel}, which are considered as the foundation for a number of techniques proposed recently. Each of those has a slightly different approach, which is summarized in Figure \ref{fig:overview}. 

At the moment, containers have slowly replaced virtual machines to become the de facto standard to deploy cloud services. At the edges, where the resources are limited and heterogeneous, it is even more favorable to deploy containerized services due to its low virtualization overhead and fast responsiveness. In certain cases, containers can even achieve near-native performance \cite{morabito2018consolidate, kovacs2017comparison}. However, containers migration in general and containerized microservices migration in particular have not been well mentioned in published works so far \cite{puliafito2019container}. Over the years, several new techniques have been developed and in the following, we mention several state-of-the-art techniques developed specifically for container live migration, namely Virtuozzo, Picocenter, ClusterHQ’s Flocker, and Voyager.

Virtuozzo \cite{mirkin2008containers} provides a bare-metal virtualization solution that supports containers and can facilitate live migration. Firstly, the container’s file system and virtual memory are transferred to the target host. After this stage, Virtuozzo freezes all container processes and disable networking. As the container is stopped, its memory state is dumped into a file, and this file is transferred to the target host. From this moment, one or several synchronization rounds take place, in which changed memory page and disk block since the last transfer are migrated. When this iterative process ends, the container is resumed at the target host. The implemented technique is based on the post-copy technique mentioned above, which can achieve a relatively short downtime and requires a small amount of data to be transferred. However, like any other post-copy migration technique, it is only feasible when the amount of changed memory pages and disk blocks (\textit{deltas}) is small, thus outage time imperceptible. Therefore, if there is a high latency connection between source and target host, as well as when the page dirtying rate is high, for example in data intensive applications, the total migration time can be too long to be acceptable.

Picocenter \cite{zhang2016picocenter} has been built with the ability to swap out inactive containers from the cloud to object store (Amazon S3) and swap in them back on demand. Under the hood, it uses Checkpoint/Restore In Userspace (CRIU) to capture memory-state and \textit{btrfs file system} to store the persistent state. In addition, \textit{ActiveSet} was proposed to allow memory pages to be on-access and lazily restored. 

Flocker by ClusterHQ is an open-source container data volume manager developed specifically for Docker containers. It supports migration for network-attached storage backends like Amazon EBS, Openstack Cinder, VMware vSphere etc., by re-attaching these network storage for containers. However, locally attached volume migration is supported only for ZFS filesystem. 

Voyager \cite{nadgowda2017voyager} is a file system-- and vendor--agnostic migration service developed for OCI (Open Container Initiative)--compliant containers. Voyager enables just-in-time live container migration with short downtime, by discovering all data endpoints of a container and migrating in-memory state via CRIU. Network attached storage is manually unmounted from the source host and mounted again at the target host. With a union view of the data between the source and target hosts, Voyager containers can resume an operation instantly on the target host, while performing disk state transfer either on-demand (Copy-on-Write) or through lazy replication. 

\subsection{Asynchronous messaging in microservices-based applications}

Considered as a new architectural style for cloud applications, microservices are designed to be self-contained, so that each service can be developed and deployed independently. To extend this independence into operation, the microservices usually rely on an external communication infrastructure for exchanging data among services. This design allows developers to focus on the actual microservices' business logic rather than on implementing mechanisms for sending and receiving data. Among others, asynchronous messaging is widely implemented for microservices, as this communication scheme can ensure that no microservice has to wait for another service's response \cite{dinh2019maia}. In this design pattern, all communication among services are encapsulated in messages, and they are handled asynchronously by a message broker, hence the name ``asynchronous messaging". This concept is also known as ``smart endpoints and dumb pipes concept" \cite{hilgert2016optimization}, in which the term ``dumb pipes" refers to simple and lightweight asynchronous messaging like Advanced Message Queuing Protocol (AMQP) or Message Queuing Telemetry Transport (MQTT), both are widely supported by a number message brokers such as RabbitMQ or ZeroMQ \cite{shadija2017towards}. 

In addition, to decouple client apps from microservices, minimize the service calls and reduce the attack surface, microservices are usually not built with public endpoints, i.e., external clients cannot consume services directly, but rather through an API gateway \cite{montesi2016circuit}. Therefore, a major proportion of communications among microservices are done internally, especially in event-driven design, where all events are published and exchanged across microservices as messages. Those messages are exchanged via publish/subscribe queues provided by the message broker. This means, in such a design, the message broker plays a crucial role in coordinating the entire cluster of microservices. Given that asynchronous messaging in general and the message broker in particular both are widely used in various microservice implementations \cite{bakshi2017microservices}, this work's main idea is to make use of those components to reproduce the states of migrated services during the migration process.

\section{Concept and Design}

Compared to the above-mentioned techniques, this work proposes a novel migration scheme named MS2M, which also utilizes CRIU to create checkpoints and restore the containerized microservice at the target host. However, the novelty of this work lies in the \textit{state synchronization mechanism}. Instead of transferring service states directly by copying in-memory data, which is proved complex by previous works, this proposal indirectly rebuilds the state of the migrated instance by replaying incoming messages during the migration period. In this section, we present in detail how this idea works and discuss several of its fundamental characteristics. An in-depth discussion based on experimental evaluation results is given in the next section.

In this proposal, we differentiate three main involving actors, which are described as below:

\begin{itemize}
    \item \textbf{Source host:} The server, where the microservice is originally hosted.
    \item \textbf{Target host:} The new server, where the microservice will be migrated to.
    \item \textbf{Migration manager:} This entity's main responsibility is to monitor and coordinate the migration process among source host, target host and the microservice. This can be deployed as another microservice within the application itself, thus enabling the self-management capability of the application.
\end{itemize}

The three actors interact with each other during the migration process, which can be divided into five main phases: Service checkpoint creation, Checkpoint transfer, Service restoration, Messages replay, and Finalization. These interactions are illustrated in Figure \ref{fig:replayqueue} while the details of each phase are explained below:

\begin{figure}
\includegraphics[width=\linewidth]{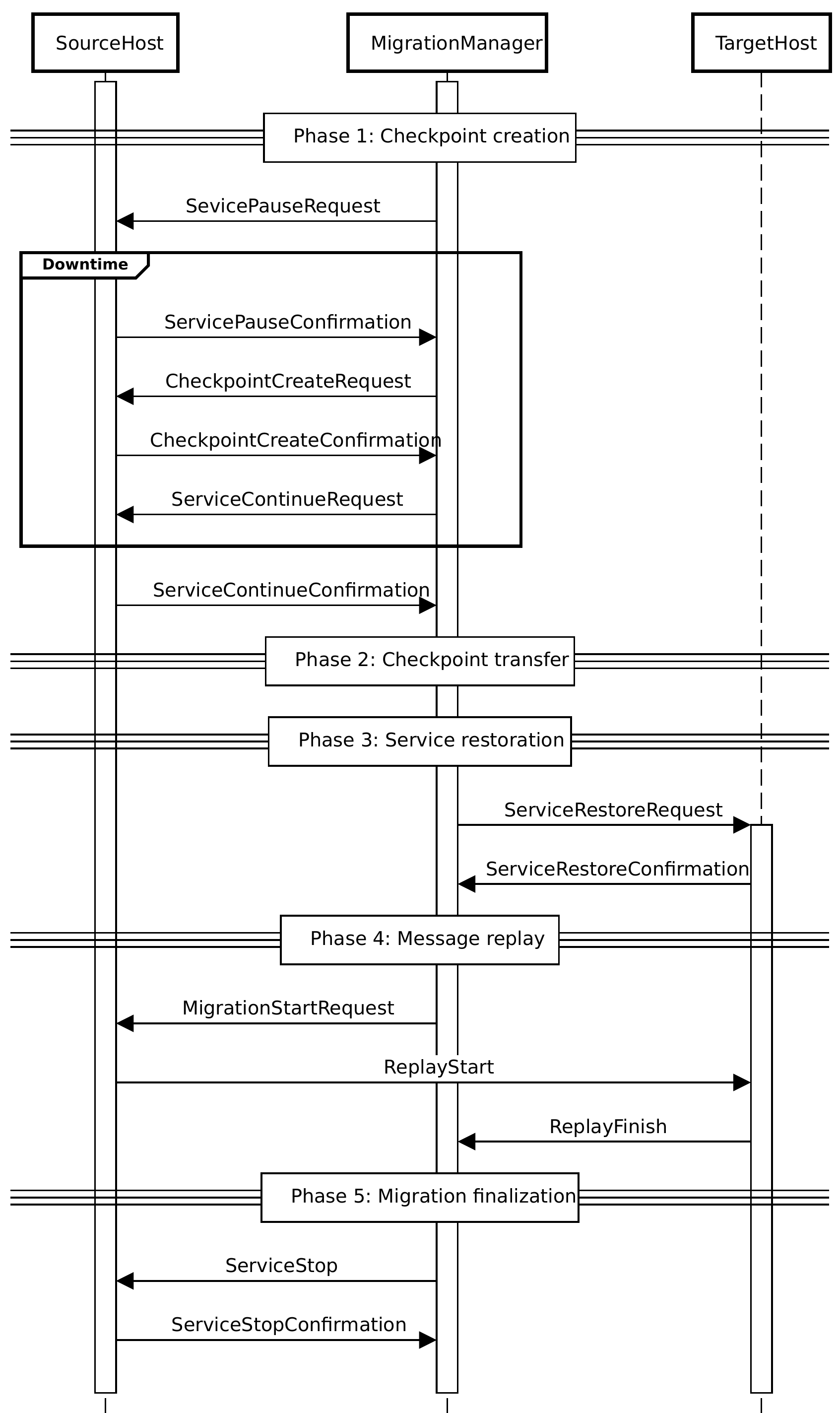}
    \caption{Detailed migration process of MS2M.}
    \label{fig:replayqueue}
\end{figure}

\begin{enumerate}
\item \textbf{Checkpoint creation:} The Migration Manager sends a \textit{ServicePauseRequest} to the Source Host to temporarily pause the service instance and prepare to create a service checkpoint. During this preparation phase, the service is unsubscribed from the \textit{main message queue} before a checkpoint is created. This is necessary to prevent the newly restored service at the Target Host from consuming the same message queue immediately after restoration. After the checkpoint is created, a \textit{secondary message queue} is generated, which keeps a copy of every message coming to the \textit{main message queue} since the migration initiation. Then, the service instance at the Source Host is resumed and continues to operate normally. Upon this phase completion, the Migration Manager is informed to proceed to the next phases.

\item \textbf{Checkpoint transfer:} During this phase, the checkpoint is transferred to the Target Host. This process happens in the background without interrupting service operation. The incoming requests are handled by the original service instance as normal.

\item \textbf{Service restoration:} After the checkpoint is fully transferred to the Target Host, the service instance is restored and configured to subscribe to the \textit{secondary message queue}. When the newly restored service instance functions properly, the Migration Manager is informed by a confirmation from the Target Host.

\item \textbf{Message replay:} In this phase, the restored service instance at the Target Host starts processing messages in the \textit{secondary message queue} without any output to synchronize the state with the original instance. Because all the message that arrives since the migration initiation are duplicated from the \textit{main message queue} to the \textit{secondary message queue}, the original service instance and the migrated service instance have the exact same input. During this message replaying phase, the original service instance at the Source Host still operates normally. By this design, the migrated service instance can synchronize the state with the original service instance while the incoming requests are handled normally by the original instance, therefore it helps to minimize the downtime. When this phase ends, the original instance notifies the migrated instance the \textit{LastMessageProcessedID}, to indicate which messages have been processed successfully. 

\item \textbf{Finalization:} After replaying the \textit{LastMessageProcessedID}, the migrated instance switches back to the \textit{main message queue}. In addition, from this point in time, the processed output from the migrated instance is sent normally to the message broker. The original instance stops and the migration process is finally completed.
\end{enumerate}

\begin{figure}
\includegraphics[width=\linewidth]{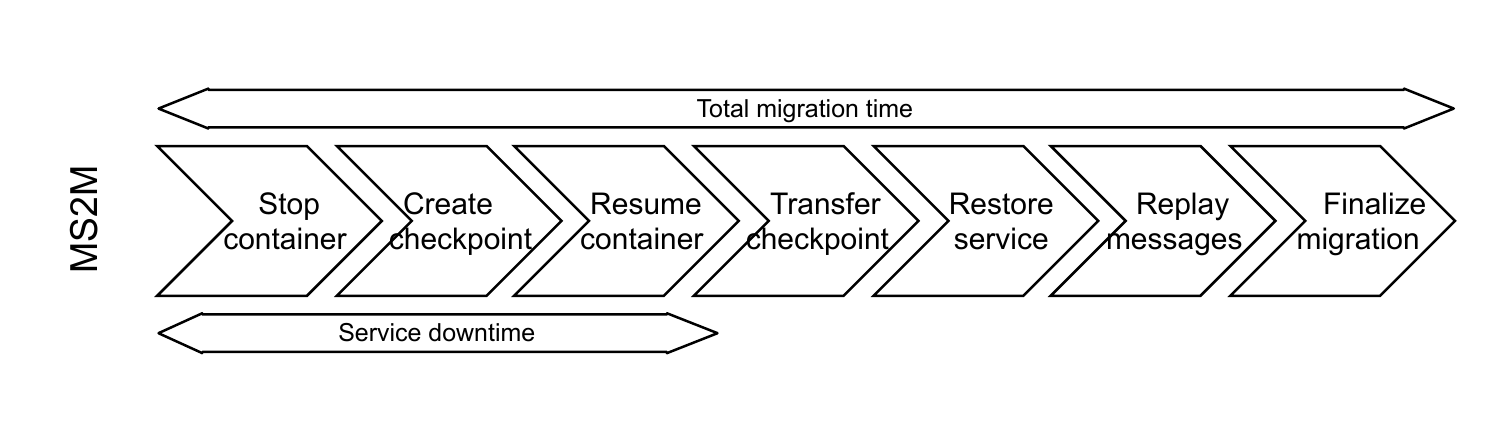}
    \caption{Overview of the proposed MS2M scheme.}
    \label{fig:ms2m}
\end{figure}

Figure \ref{fig:ms2m} summarizes the above process, showing the Total migration time and the Service downtime of MS2M. Compared to the similarly illustrated techniques in Figure \ref{fig:overview}, one can immediately recognize the key difference: The dump transfer process in other techniques is included in the service downtime, while in MS2M, the checkpoint transfer phase is not, which can potentially lead to significant shorter downtime. This is made possible because in MS2M, the state difference is tracked indirectly by capturing messages to the \textit{secondary message queue}. Therefore, the original service instance can be resumed immediately after checkpoint creation, and the checkpoint can be transferred to the Target Host in the background. The state difference occurs during the period between checkpointing and service instance restoration at the Target Host can be easily re-synchronized by replaying messages from the \textit{secondary message queue}. During this process, the original instance still processes messages normally, and this task will only be handed over to the new instance when the message replay phase ends, i.e., when the service states in the original instance and the new instance are synchronized.

In addition to shorter service downtime, compared to previously discussed approaches, this proposed scheme has several additional advantages:

Firstly, this is a \textit{communication--flow--aware approach} for service migration. Rather than focusing on transferring in-memory data, which is proved to be complex in previous works, MS2M employs the message broker to support the state synchronization process indirectly. By doing that, the application itself can take control over the decision when and where to migrate a particular microservice to maximize the performance. This means during the development phase, the developers based on the application’s specifics can already decide what is the optimal operation conditions for the microservices. 

Secondly, the downtime is minimized to the time required to prepare the service for checkpoint and the checkpoint process itself. It means the downtime does not depend on the network utilization, which is hard to predict during peak network traffic hours.

Lastly, in this concept, we took CRIU as an example for a container checkpoint/restore engine. However, there are several other tools that provide similar functionalities for other deployment strategies. Depending on the particular scenario, software architects can employ the most appropriate tool. The heart of this migration scheme is the coordination among microservices hosts and the replaying phase, which are both performed by the Migration Manager. This means, the exact approach can be applied for microservices--as--containers,  microservices--as--virtual machines and other deployment options.

However, this proposal also comes with certain drawbacks. One of the major ones is the unpredictable nature of the replaying process. In case the incoming message rate is higher than the processing rate of the microservices during the replay phase, this iteration could theoretically never end. This is not desirable because during this phase, at least double the resources are occupied for both service instances to process the same request twice. On top of that, both the Migration Manager and the additional queue create overhead in the communication channel. Another possible issue is that this migration scheme doesn't guarantee the completion of migration if the original instance stops working during the process. Last but not least, there might be a degraded service period after the service instance finishes replaying messages, when the message processing latency is higher than normal.

In this section, we have presented a detailed description of the proposed migration process together with some preliminary analysis on its advantages and disadvantages. In the next sections, with the help of an interactive game prototype, we provide a more thorough understanding of the proposed migration techniques across several criteria such as latency performance.

\section{Experimental evaluation}

\subsection{Experimental Setup}

\begin{figure}
\includegraphics[width=\linewidth]{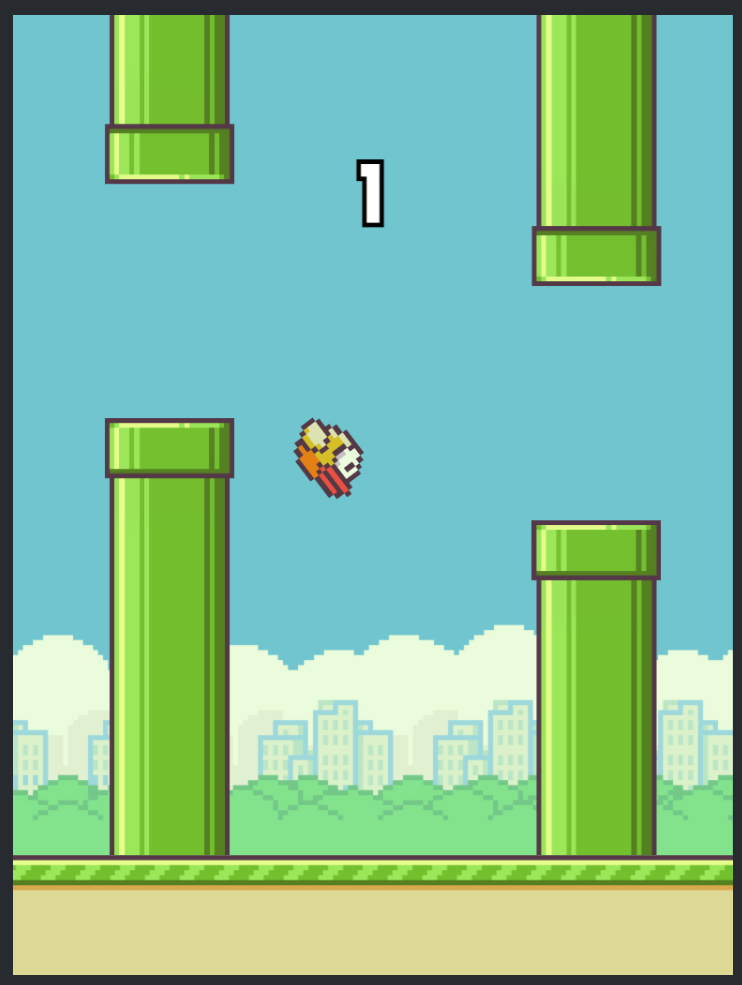}
    \caption{The player's mission is to keep the bird flying without hitting the pipes.}
    \label{fig:flappybird}
\end{figure}

For demonstration and evaluation purposes, we implement an interactive game prototype. This is a remake version of an iconic mobile game named Flappy Bird. In this arcade-style game, there is a bird that moves persistently to the right and automatically descends. The player should constantly enter inputs via the keyboard to keep the bird flying and avoid the pipes as shown in Figure \ref{fig:flappybird}. Whenever the player creates an input using the space bar, the bird ascends. The score is calculated based on the number of pipes that the bird can safely pass. 

To evaluate our MS2M migration technique, a new cloud-based microservice has been added to persist state data from the game session, including players' game settings and scores. While the user interacts with the browser-based component, the microservice is migrated to the cloud region with better latency performance and related data is recorded via various tracers. The detailed evaluation of these data is presented later in the second part of this section.

The whole application are deployed as four virtual machines on Google Cloud Platform (GCP) with the following configurations:

\begin{enumerate}
    \item \textbf{Broker:} configured using GCP's e2-medium (2 vCPUs, 4 GB memory).
    \item \textbf{Web App:}  configured using GCP's e2-micro (2 vCPUs, 1 GB memory).
    \item \textbf{Source and Target host: } Two virtual machines located in Belgium and the USA to host the containerized microservice. Both are configured using e2-medium (2 vCPUs, 4 GB memory)
\end{enumerate}

All virtual machines are deployed with Ubuntu 20.10 Groovy. The setup uses Podman 3.1.0 as container engine and RabbitMQ 3.8.11 as message broker. The Migration Manager is deployed as a microservice, and it communicates with the host via Podman's API. This API supports executing CRIU's command directly, such as \textit{create checkpoint}, \textit{start container}, \textit{stop container}, \textit{resume container}. Regarding the virtual machines' locations, only one of them (Source Host) is located in the USA, while all other machines are hosted in Belgium. 

The evaluation test is performed by having a person playing the game on a computer. During this game session, a script is run to automatically migrate the microservices back and forth between GCP's computing region in the US and Belgium. We ran in total 50 migration tests, but because the network topology and load on GCP cluster might affect the evaluation results, we only consider the migrations in one direction from the USA to Belgium (in total 25 data points). 

\subsection{Evaluation results}

As in any migration technique, the delay performance is one of the most crucial evaluation criteria. Similar to many other previous works when evaluating a certain technique, we evaluate two important values, which are the \textit{Total migration time} and \textit{Service downtime}. 

The \textit{total migration time} is counted from the initiation of the migration until the service is fully migrated and works properly at the new location, while the \textit{service downtime} corresponds to the duration in which the service is completely unavailable. Long service downtime is obviously not favorable, as more user requests will be rejected in this case. From users' perspectives, the total migration time is less significant, however, a long total migration time results in more resource utilization and a longer service degradation period. This means, in the scenario where multiple simultaneous migration processes take place, a long migration time can lead to excessive resource usage. Therefore, an ideal migration technique should be able to minimize both of those values. 

Based on those definitions, in this particular example, these periods are defined as follows:

\begin{enumerate}
    \item The \textit{total migration time} starts when the Migration Manager initiates the migration process by sending the \textit{ServicePauseRequest} to the Source Host. This period ends with the stop of the original service instance at the Source Host while the service is resumed by the new service instance at the Target Host. 
    \item The \textit{service downtime} in this scenario equals to the Service checkpoint creation time, in which the original service instance is offline and there is no service instance serving the incoming request. However, it is important to note that due to the asynchronous nature of the message broker, no user request is rejected, thus the user perceives this downtime as a degraded service period when it takes longer as usual to process requests.
\end{enumerate}

To evaluate our proposed MS2M, in the prototype we also implemented the stop--and--copy migration technique as the baseline for comparison. The details of stop-and-copy migration technique implemented in this test are summarized in Figure \ref{fig:stopcopy} below. 
\begin{figure}
\includegraphics[width=\linewidth]{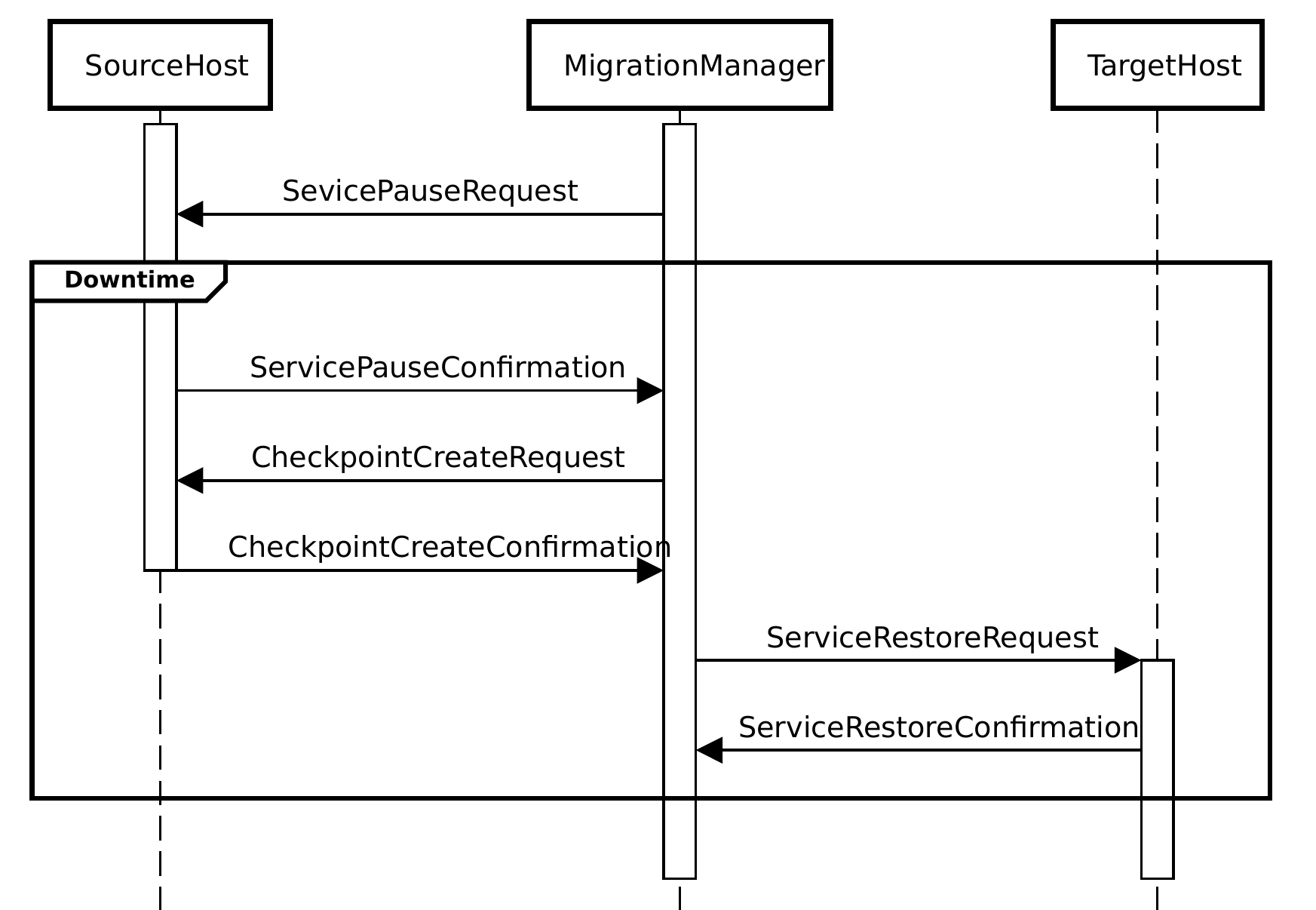}
    \caption{Stop-and-copy process.}
    \label{fig:stopcopy}
\end{figure}

\begin{figure}
\includegraphics[width=\linewidth]{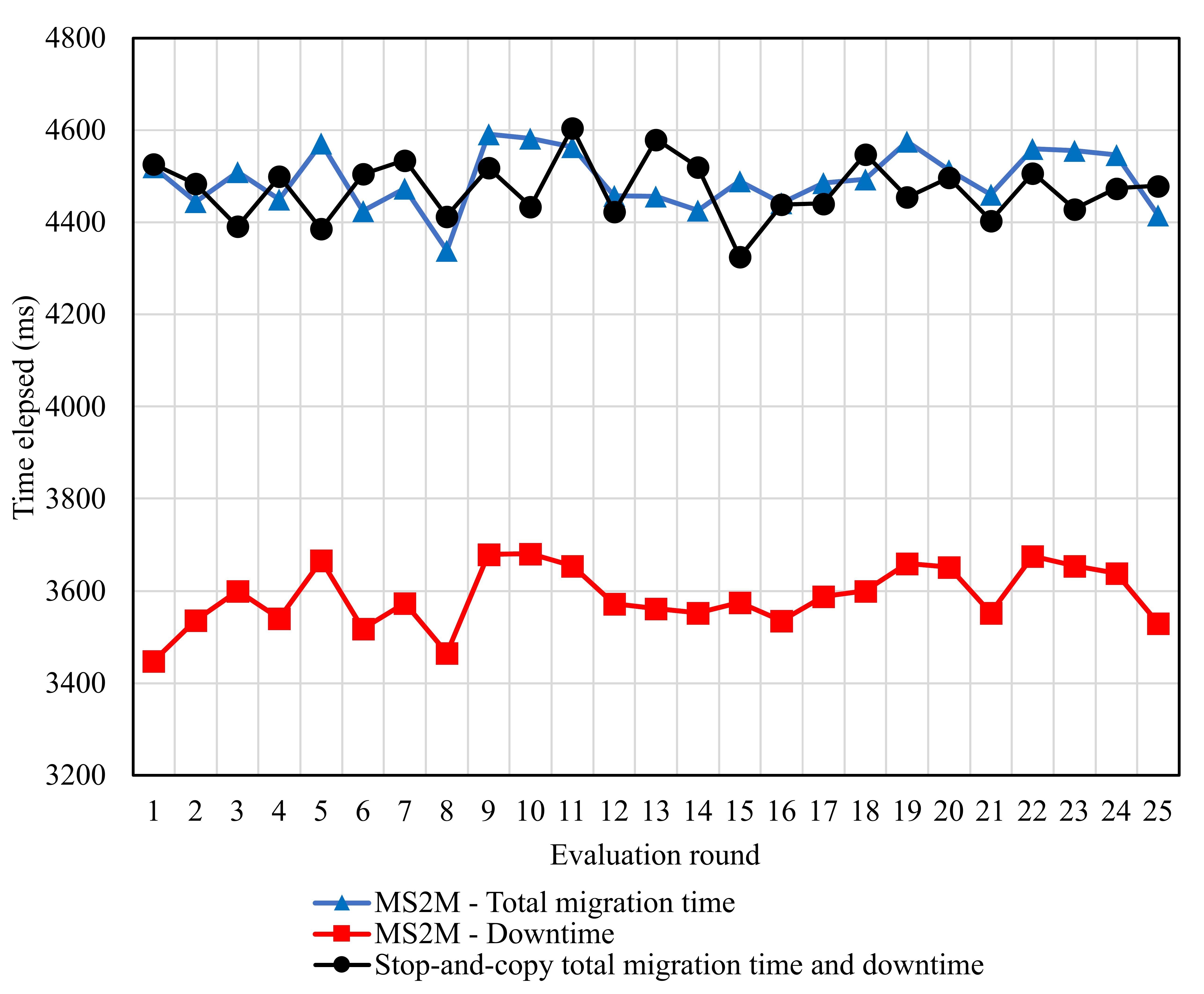}
    \caption{Overall time performance comparison between stop-copy method and the proposed method.}
    \label{fig:overallresults}
\end{figure}

\begin{figure}
\includegraphics[width=\linewidth]{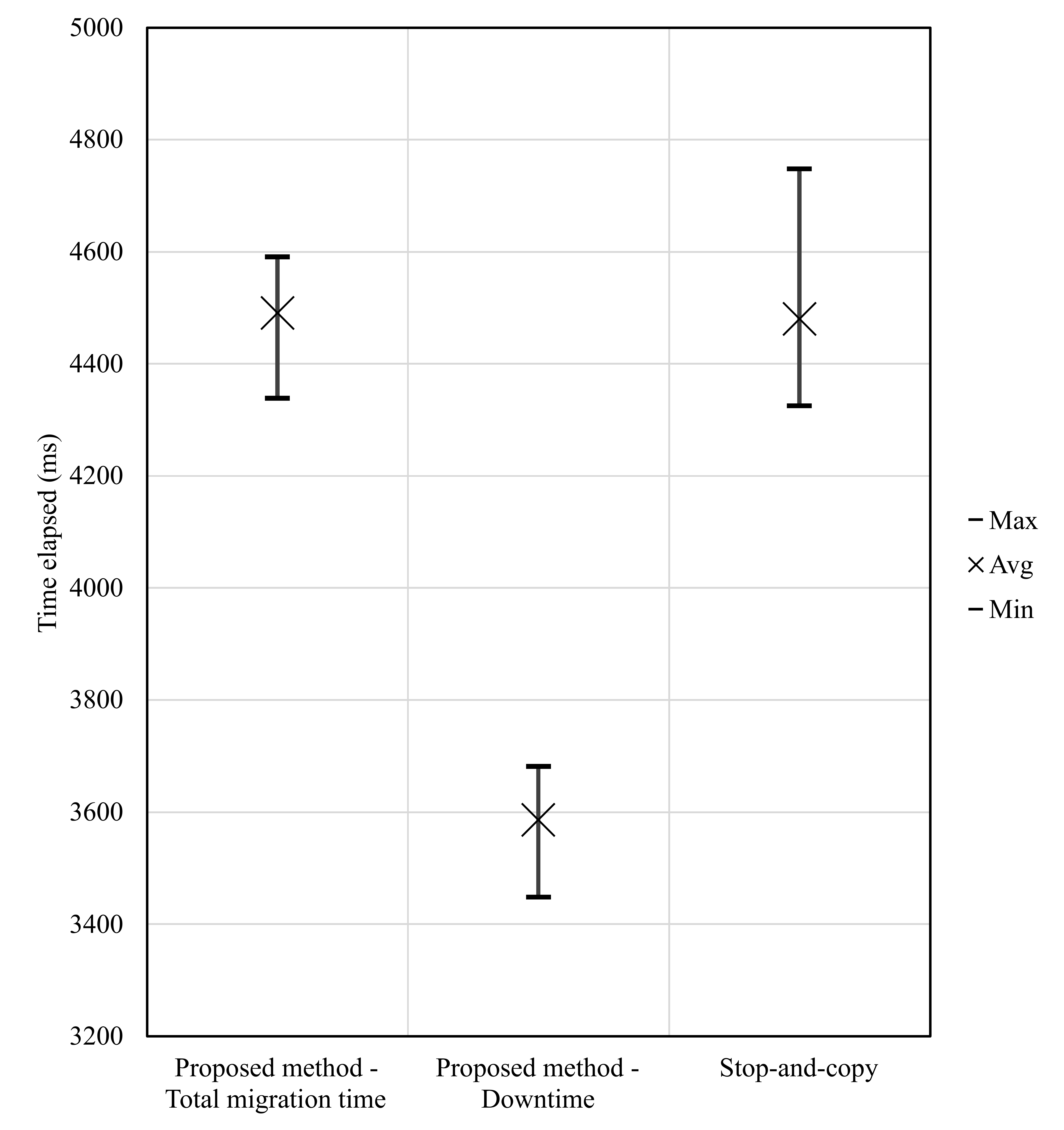}
    \caption{Overall time performance comparison between stop-copy method and the proposed method.}
    \label{fig:overallresultsboxallresultsbox}
\end{figure}

For the evaluation, we collect data from 25 evaluation iterations, each iteration is done by both migration techniques implemented. The detailed results are presented in Figure \ref{fig:overallresults} and Figure \ref{fig:overallresultsboxallresultsbox}. 

The results show a similar total migration time in both schemes. While the stop-and-copy migration needs 4472.72ms, the proposed method takes slightly longer -- 4852.86ms on average (±8.50\% more) -- to complete the migration process. Interestingly, the evaluation confirms our expectation of a shorter downtime by MS2M: The new scheme has only 3581.84ms downtime, yields a 19.92\% reduction from the stop--and--copy scheme. 

\begin{figure}
\includegraphics[width=\linewidth]{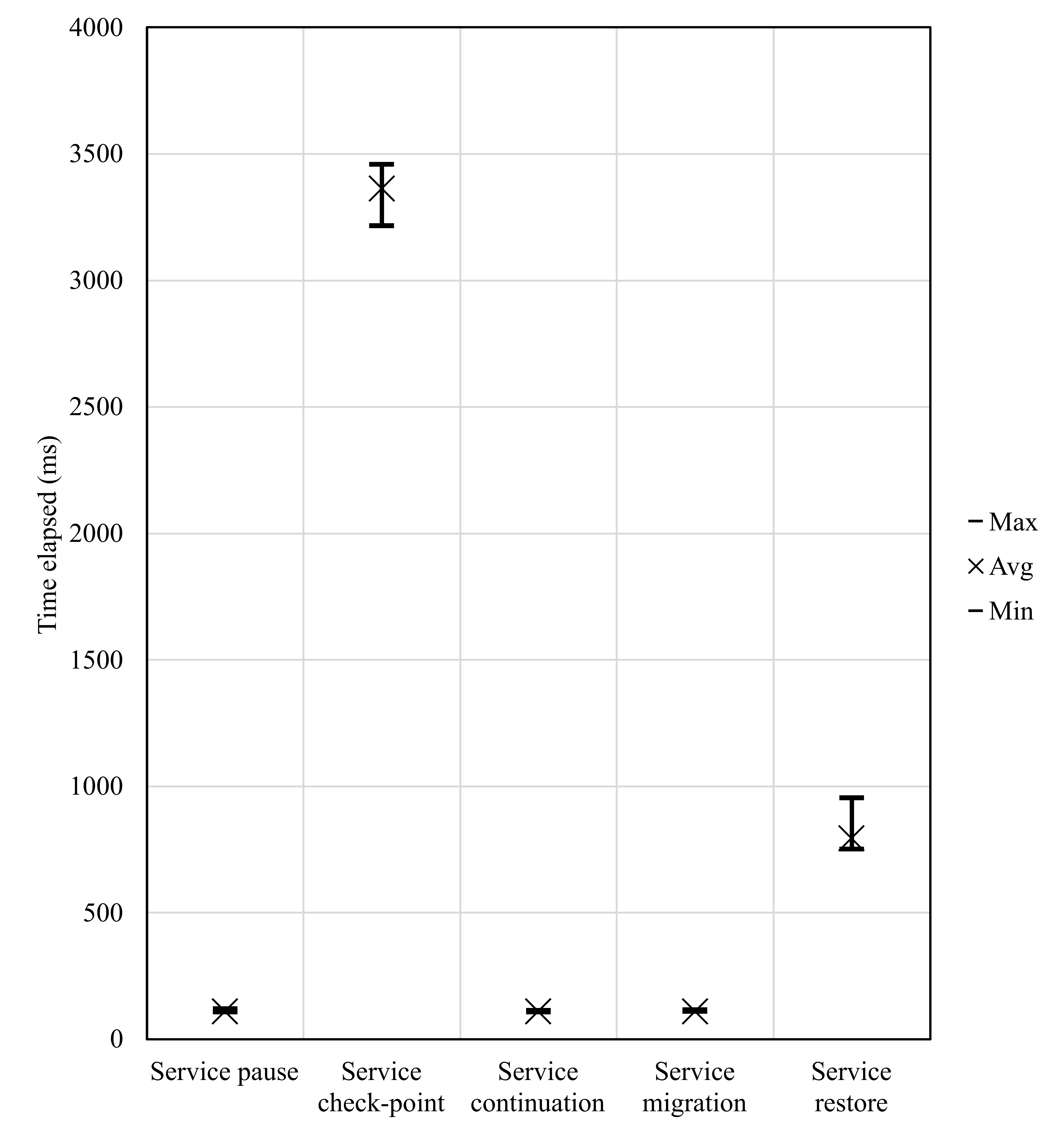}
    \caption{Time performance analysis of the proposed MS2M method.}
    \label{fig:overallresultsphases}
\end{figure}

To gain a further understanding of the delay performance of the proposed technique, we also performed a more detailed time analysis to reveal the distribution of elapsed time in different phases as shown in Figure \ref{fig:overallresultsphases}. Data for five main phases are collected, which are \textit{Service pause} (time elapsed to pause the original instance), \textit{Service checkpoint} (time elapsed to create the instance's checkpoint), \textit{Service continuation} (time elapsed to continue the original instance after checkpoint), \textit{Service migration} (time elapsed to migrate the checkpoint), and \textit{Service restoration} (time elapsed to restore the instance at Target host). While \textit{Service pause}, \textit{Service continuation}, and \textit{Service migration} account for negligible proportions; i.e. 1.23\%, 1.23\%, and 9.54\% respectively, the Service checkpoint and Service restoration process account for 63.04\% and 24.97\% of the total migration time. 

These experimental results complement the advantages presented in Section III, demonstrating the feasibility of MS2M to use the message queues to support the migration process. By using an additional message queue, this setup can reduce the downtime by nearly 20\%. This means that fewer client requests arrive without being processed. In addition, this newly proposed scheme can maintain a comparable total migration time with only 8.50\% difference compared to the stop--and--copy method. These initial results can be used to prove the feasibility of our proposal in a cloud-based microservices environment.

However, there are also some important remarks. Firstly, although the experimental results show a comparable total migration time between the proposed scheme and the stop-and-copy method, it is necessary to note that in case the message arrival rate is high, the results might look entirely different. This is similar to the \textit{high page dirty rate} scenario in the post-copy scheme: The higher this rate is, the longer it takes to finish the migration. Especially when the message arrival rate is higher than the message processing rate, it could theoretically take infinite time to finish the message replay phase, resulting in an unacceptable long total migration time. As mentioned before, this could become critical as more resources are utilized, hence the need to minimize this period. 

In addition, the checkpoint and restoration process accounts for a major proportion of the migration time (88.04\%). As pointed out by previous work, the checkpoint time increases linearly with the size of applications' memory state \cite{nadgowda2017voyager}. This suggests microservices with a larger memory footprint can also reduce the performance of this migration technique.

In this particular use case of an interactive game, both the amount of incoming data and its rate to the microservice are not significantly high. This explains why the total migration time of the two migration techniques recorded in this evaluation are not much different. Therefore, it should be noted that depending on the nature of the application, the total migration time could be longer compared to stop-and-copy scheme. 

\section{Conclusion}

An ideal service-oriented microservices-based application should be developed as stateless to separate data management responsibilities from services, thus reducing resource consumption and improving the overall scalability. However, we have pointed out that this is not always feasible, i.e., stateful microservices are essential in certain applications. However, dealing with those stateful services requires special attention to those state data, which is reflected in the case of service migration. This paper presents the design, implementation, and evaluation of a novel approach for live stateful microservice migration. It shows how the asynchronous message broker can be utilized to support the migration process and enable service migration without the need to directly handle in-memory data. By using the message broker to spawn a new message queue and indirectly track the state difference during migration, this method reduces the downtime by as much as 19.92\% compared to the stop-and-copy method. Although the prototype used in this work is deployed as containers, the concept of using the message broker to support the restoration of state information can be generalized to apply to any event-driven microservices regardless of deployment strategy.

However, there are still some characteristics of the message replay process, which we need to pay special attention to. Firstly, if the incoming message rate is high, especially in the extreme case, when this value is higher than the processing rate, it could take a long time to finish the migration process, leading to excessive resource usage. In addition, services with a large memory footprint could also reduce the performance of this migration technique, because the service checkpoint and restore process largely depends on the service size, and they are both main contributors to the total migration time.

The results taken from this work lead to several possible future research directions to address open questions. An in-depth comparison with more recent migration techniques offered by cloud platforms such as Kubernetes would bring new insights, especially in real world scenarios. We also plan to further investigate possible solutions to optimize the checkpoint and restore processes, for example with an \textit{incremental checkpoint} technique. In addition, the time performance of this proposed scheme needs more extensive analysis, especially the message replay process. Another use case with a high frequency of messages exchanged among various service types might help to accurately characterize this method's performance.

\section{Acknowledgement}

We would like to show our gratitude to our students Xhorxhina Taraj, Spoorthi Satheesha, Tuan Anh Tran, Zubair Hossain, Sanjeet Raj Pandey, and Dimitrios  Peppas for assisting us during the course of this research.

\bibliographystyle{IEEEtran}
\bibliography{bibliography}
\end{document}